\documentclass[%
 aip,
 amsmath,amssymb,
 reprint,%
]{revtex4-1}

\usepackage{graphicx}
\usepackage{dcolumn}
\usepackage{bm}

\usepackage[utf8]{inputenc}
\usepackage[T1]{fontenc}
\usepackage{mathptmx}
\usepackage{etoolbox}
\usepackage{xcolor}
\usepackage{svg}

\newcommand{\SIQSE}{\affiliation{1}{Shenzhen Institute for Quantum Science and Engineering, Southern University of Science and Technology, Shenzhen, Guangdong, China}}
\newcommand{\DPHY}{\affiliation{2}{Department of Physics, Southern University of Science and Technology, Shenzhen, Guangdong, China}}
\newcommand{\IQA}{\affiliation{3}{International Quantum Academy, Shenzhen, Guangdong, China}}
\newcommand{\GDKL}{\affiliation{4}{Guangdong Provincial Key Laboratory of Quantum Science and Engineering, Southern University of Science and Technology, Shenzhen, Guangdong, China}}
\newcommand{\HFNL}{\affiliation{5}{
Shenzhen Branch, Hefei National Laboratory, Shenzhen 518048, China}}

\begin{document}


\title{\textit{In situ} mixer calibration for superconducting quantum circuits}
\author{Nan Wu}
\affiliation{\SIQSE}\affiliation{\IQA}\affiliation{\GDKL}
\author{Jing Lin}
\affiliation{\SIQSE}\affiliation{\IQA}\affiliation{\GDKL}
\author{Changrong Xie}
\affiliation{\SIQSE}\affiliation{\IQA}\affiliation{\GDKL}
\author{Zechen Guo}
\affiliation{\SIQSE}\affiliation{\IQA}\affiliation{\GDKL}
\author{Wenhui Huang}
\affiliation{\SIQSE}\affiliation{\IQA}\affiliation{\GDKL}
\author{Libo Zhang}
\affiliation{\SIQSE}\affiliation{\IQA}\affiliation{\GDKL}
\author{Yuxuan Zhou}
\affiliation{\IQA}
\author{Xuandong Sun}
\affiliation{\SIQSE}\affiliation{\IQA}\affiliation{\GDKL}\affiliation{\DPHY}
\author{Jiawei Zhang}
\affiliation{\SIQSE}\affiliation{\IQA}\affiliation{\GDKL}
\author{Weijie Guo}
\affiliation{\IQA}
\author{Xiayu Linpeng}
\affiliation{\IQA}
\author{Song Liu}
\affiliation{\SIQSE}\affiliation{\IQA}\affiliation{\GDKL}\affiliation{\HFNL}
\author{Yang Liu}
\affiliation{\IQA}
\author{Wenhui Ren}
\affiliation{\IQA}
\author{Ziyu Tao}
\affiliation{\IQA}
\author{Ji Jiang}
\email{jiangj3@sustech.edu.cn}
\affiliation{\SIQSE}\affiliation{\IQA}\affiliation{\GDKL}
\author{Ji Chu}
\affiliation{\IQA}
\author{Jingjing Niu}
\affiliation{\IQA}\affiliation{\HFNL}
\author{Youpeng Zhong}
\affiliation{\SIQSE}\affiliation{\IQA}\affiliation{\GDKL}\affiliation{\HFNL}
\author{Dapeng Yu}
\affiliation{\SIQSE}\affiliation{\IQA}\affiliation{\GDKL}\affiliation{\DPHY}\affiliation{\HFNL}

\date{\today}

\begin{abstract}
Mixers play a crucial role in superconducting quantum computing, primarily by facilitating frequency conversion of signals to enable precise control and readout of quantum states. However, imperfections, particularly carrier leakage and unwanted sideband signal, can significantly compromise control fidelity. To mitigate these defects, regular and precise mixer calibrations are indispensable, yet they pose a formidable challenge in large-scale quantum control. Here, we introduce an \textit{in situ} calibration technique and outcome-focused mixer calibration scheme using superconducting qubits. Our method leverages the qubit's response to imperfect signals, allowing for calibration without modifying the wiring configuration. We experimentally validate the efficacy of this technique by benchmarking single-qubit gate fidelity and qubit coherence time.
\end{abstract}

\maketitle

Superconducting circuits have emerged as one of the most promising platforms for large-scale, fault-tolerant quantum computing~\cite{marques2022logical,google2023code,ibm2023127q,cao2023generation}, spurring active research in various applications, including quantum simulation, quantum machine learning, and quantum algorithms~\cite{google2020hartree,mi2022time,xiang2023simulating,PhysRevLett.131.080401,PhysRevLett.122.010501,yao2023observation,xu2024non,tao2023interaction,PhysRevLett.132.020601,google_supremacy,pan2023deep,guo2024experimental,chu2023scalable}. The operational frequencies of superconducting qubits and resonators utilized for readout typically reside within the 4-10 GHz range~\cite{koch2007charge}. To attain high-fidelity control and measurement of these superconducting qubits, it is imperative to employ precise microwave pulse sequences with nanosecond-level timing accuracy~\cite{sheldon2016procedure,li2023error,krinner2020demonstration,ni2022scalable,liu2023single,jiang2024situ,yxh2024coupler}. Typically, control pulses are initially generated at an intermediate frequency (IF) by arbitrary waveform generators (AWGs) and then up-converted to microwave frequencies through mixing with a carrier signal generated by a local oscillator (LO). Mixers, especially in-phase and quadrature (IQ) mixers, play a crucial role in this process~\cite{ryan2017hardware,9318753,engineerguide,gao2021practical}. However, owing to their analog nature, IQ mixers require meticulous calibration to prevent the generation of spurious signals that could adversely affect control fidelity~\cite{lazuar2023calibration,you2024phase,werninghaus2021leakage}. They also suffer from substantial drift of calibration parameters over time, potentially attributed to variations in the mixer's temperature~\cite{jolin2020calibration}, necessitating regular calibration.

Conventional technique for mitigating unwanted signals is well-established at room temperature, primarily relying on spectrum analyzers for output diagnosis. Nevertheless, this method poses a significant challenge for large-scale quantum control systems, which may comprise hundreds or even thousands of channels~\cite{krinner2019engineering}, as maintaining optimal performance across all channels becomes increasingly burdensome. Although some techniques to enhance the flexibility of the conventional method have been discussed~\cite{jolin2020calibration,xu2021radio}, modifying hardware wiring can inadvertently alter the electromagnetic environment, leading to unexpected changes in quantum device parameters. Therefore, an \textit{in situ} and scalable mixer calibration scheme becomes imperative as superconducting quantum processors continue to grow in scale and complexity. On the other hand, alternative approaches that eschew the IQ mixing scheme for microwave pulse generation have been suggested, such as direct digital synthesis~\cite{raftery2017direct,PhysRevResearch.6.013305}, higher Nyquist zone microwave synthesis~\cite{kalfus2020high}, and double frequency conversion~\cite{herrmann2022frequency}.
Nonetheless, the IQ mixing scheme remains an appealing choice for large-scale quantum systems due to its simplicity, cost-effectiveness, and small feedback latency~\cite{han2023active}, despite the calibration challenges it presents.

In this Letter, we introduce and demonstrate an \textit{in situ} IQ mixer calibration method utilizing superconducting qubits. This approach capitalizes on the qubits' sensitivity to LO leakage and mirror sideband signals, eliminating the need for additional electronic devices for output diagnosis or alternation to the wiring configuration. Our method enables calibration of mixers in both the qubit's drive line and measurement line, leveraging distinct qubit responses. Furthermore, we showcase the robustness of our scheme and present an automated calibration process based on a center-searching algorithm. To validate our calibration method, we perform single-qubit gate benchmarking and monitor qubit coherence time, demonstrating that our approach effectively eliminates the detrimental effects of undesirable signals on qubit performance.

\begin{figure}
\includegraphics[width=0.5\textwidth]{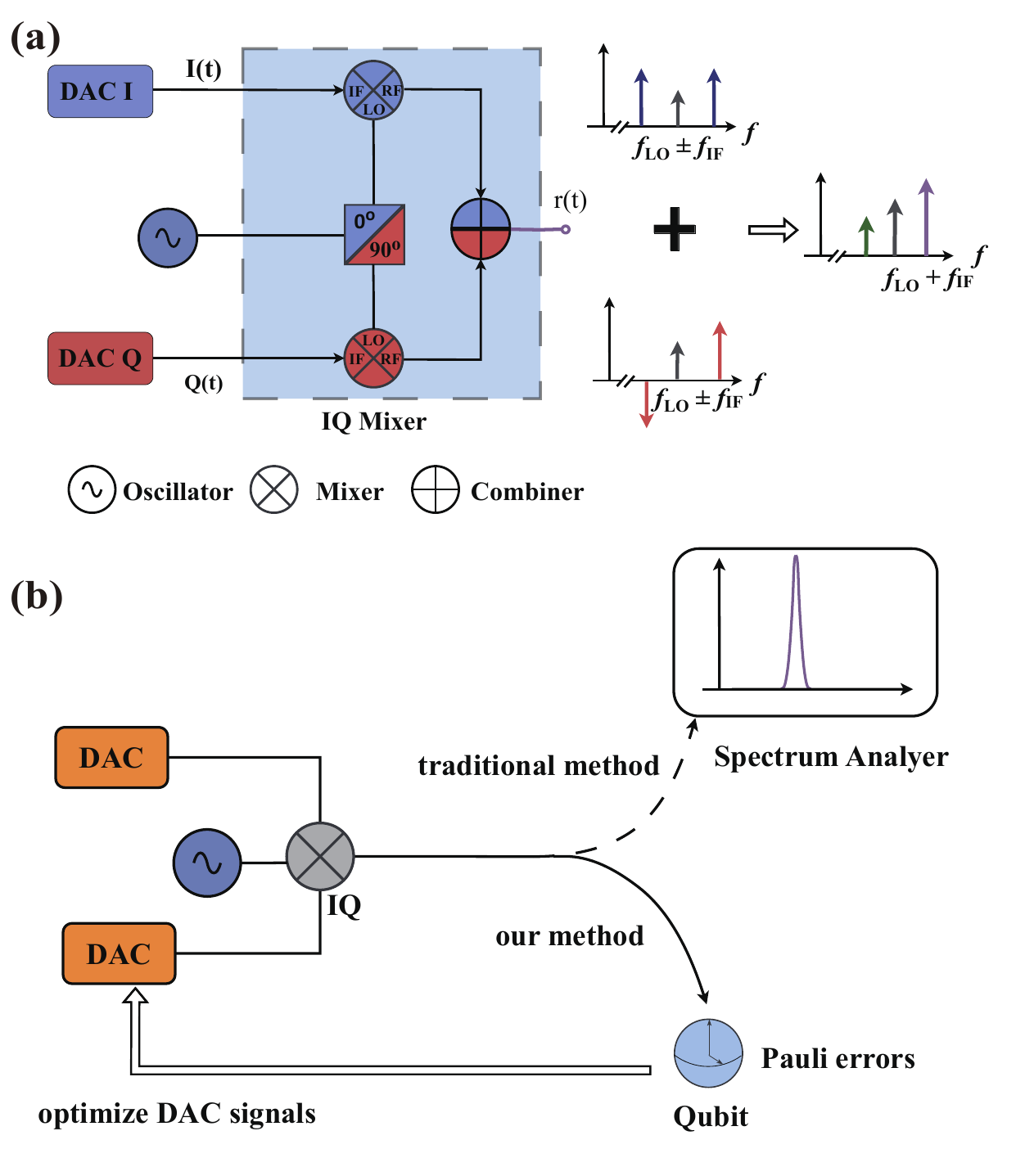}
\caption{\label{fig:background} (a) Illustration of the frequency up-conversion stage. Ideally, the output RF signal generated by two independent mixers contains two sidebands, one of which is destructively eliminated after they are superimposed in the microwave combiner. Inadequate LO-RF isolation causes the LO leakage (grey) at frequency $f_{\rm LO}$, and amplitude or phase imbalance between the I and Q ports causes the mirror sideband signal (green) at the image frequency $f_{\rm LO} - f_{\rm IF}$. (b) General concept of the \textit{in situ} mixer calibration scheme. Compared to conventional methods, our approach employs the qubit itself as a detector, preserving the integrity of the wiring configuration.}
\end{figure}

An IQ mixer consists of two single-sideband mixers operating with $90^{\circ}$ phase-shifted LO signals. It combines high-frequency signals $f_{\rm LO}$ supplied by a microwave source (LO port) and IF signals $f_{\rm IF}$ usually supplied by the digital-to-analog converters (DACs), as illustrated in Fig.~\ref{fig:background}(a). Ideally, the inherent symmetric architecture of IQ mixer ensures the output spectrum is a pure tone at the required frequency. However, LO leakage and mirror sideband signals are often encountered due to hardware imperfections. In our experiments, we modify the DAC signals fed into the I and Q ports, which are respectively 
\begin{align}
    I(t) &= A(t) \Re [e^{-2 \pi \imath f_{\rm IF} t} + c e^{2 \pi \imath f_{\rm IF} t}] + I_0, \\
    Q(t) &= A(t) \Im [e^{-2 \pi \imath f_{\rm IF} t} + c e^{2 \pi \imath f_{\rm IF} t}] + Q_0,
\end{align}
where we assume that the IF signals share the same pulse envelope $A(t)$. LO leakage can be suppressed by applying DC offsets $I_0, Q_0$ at the corresponding IF ports and the mirror sideband is compensated by adding the correction modulation signal $ce^{2 \pi \imath f_{\rm IF} t}$ characterized by a complex correction parameter $c$. The resulting radio-frequency (RF) signal is
\begin{align}
    r(t)& = A(t) \cos[2 \pi (f_{\rm LO} + f_{\rm IF})t] \label{eq:target} \\ 
    &+ A(t) \{\Re[c] \cos[2 \pi (f_{\rm LO} - f_{\rm IF})t] - \Im[c] \sin[2 \pi (f_{\rm LO} - f_{\rm IF})t]\} \label{eq:sideband} \\ 
    &+ I_0 \cos(2 \pi f_{\rm LO} t) - Q_0 \sin(2 \pi f_{\rm LO} t) \label{eq:LO_leakage},
\end{align}
The output spectrum exhibits the target frequency signal (Eq.~\ref{eq:target}), the signal for suppressing mirror sideband (Eq.~\ref{eq:sideband}) and the signal for suppressing LO leakage (Eq.~\ref{eq:LO_leakage}). Note that $|c|$ determines the amplitude of the sideband-correction signal. Instead of using a spectrum analyzer for output diagnosis, we employ the qubit itself as a detector, as depicted in Fig.~\ref{fig:background}(b). This approach optimizes the DC offsets $I_0$, $Q_0$ and the correction parameter $c$ of DAC signals by minimizing qubit's Pauli errors\cite{PhysRevB.86.180504, PhysRevLett.109.153601,sankthesis,sank2024system,PhysRevLett.116.020501} specific to the drive or the measurement line, effectively suppressing unwanted signals.

\begin{figure*}
\includegraphics[width=1.0\textwidth]{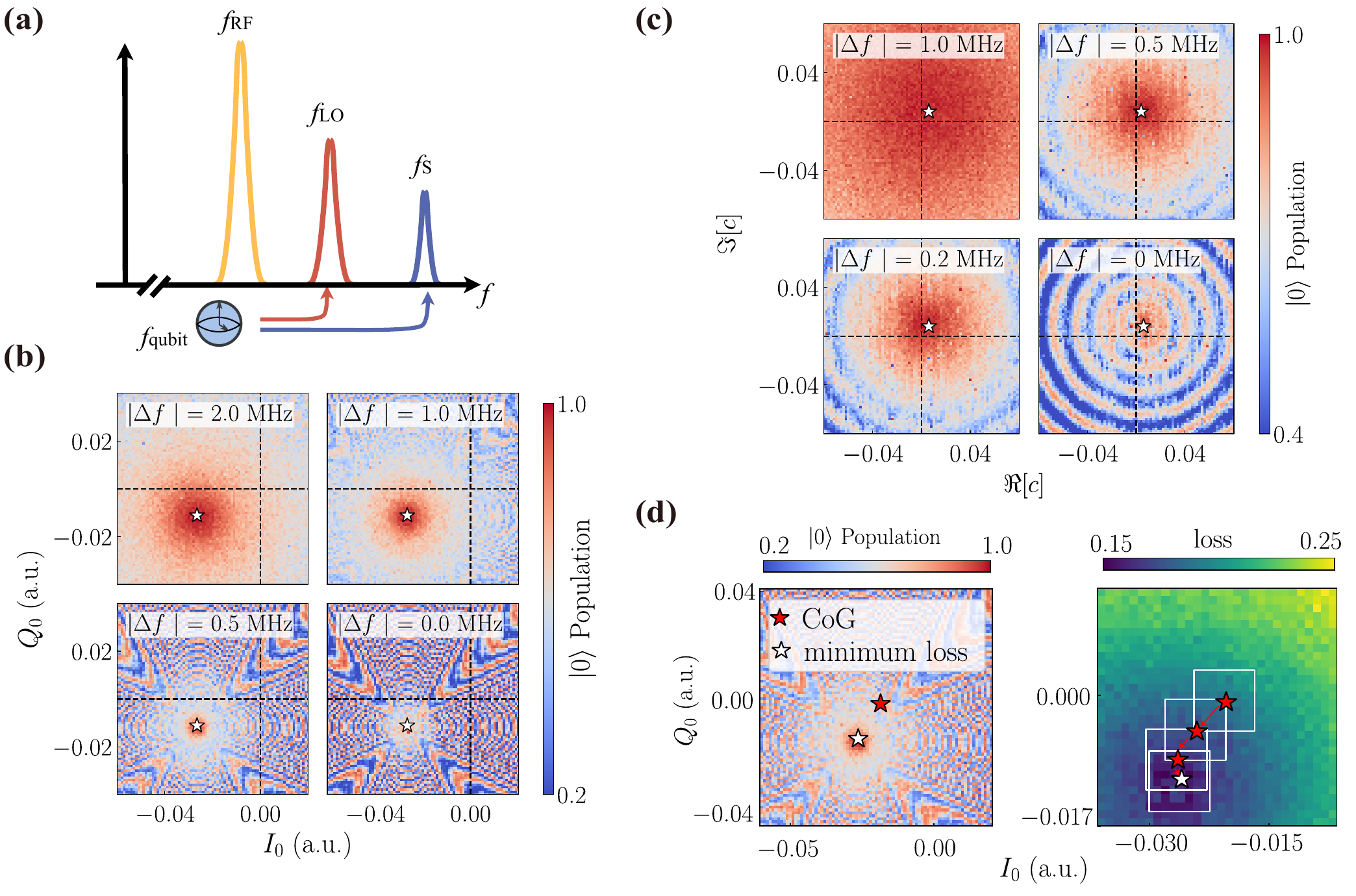}
\caption{\label{fig:dline} (a) Abstract concept of calibrating the drive mixer with a frequency-tunable qubit. Signal of LO leakage (red) or mirror sideband (blue) excites the qubit (represented by the Bloch sphere) when the qubit is biased close to the corresponding frequency. (b) $|0\rangle$ population at different LO-qubit detuning $\Delta f = f_{\rm LO} - f_{qubit}$ as a function of DC offset parameters. Results for $\Delta f > 0$ and $\Delta f < 0$ share the same tendency. (c) Ground state population at different sideband-qubit detuning as a function of the mirror sideband correction parameter $c$. (d) Visualization of the center-searching algorithm. The final result (indicated by the white star) is given by iteratively updating CoG (indicated by the red stars). The iteration steps are visualized on the landscape of the loss function in the right panel, where the white squares indicate the search range of the four iteration steps.}
\end{figure*}

The IQ mixer of the drive line (drive mixer) is calibrated by minimizing qubit excitation ($|1\rangle$ population) induced by unwanted signals or equivalently, maximizing the ground state ($|0\rangle$) population. As shown in Fig.~\ref{fig:dline}(a), the qubit frequency $f_{qubit}$ is tuned close to either LO frequency $f_{\rm LO}$ or the mirror sideband frequency $f_{S}=f_{\rm LO}+f_{\rm IF}$. According to Eq.~\ref{eq:target}-\ref{eq:LO_leakage}, the output RF signal exhibits only the compensated leakage signal when $A(t)=0$. The LO leakage can be suppressed with proper offset parameters $I_0$ and $Q_0$. Fig.~\ref{fig:dline}(b) shows leakage-induced excitation patterns versus the offset parameters. As the intensity of the leakage-correction signal is proportional to $\sqrt{I_0^2+Q_0^2}$, the excitation pattern is centrosymmetric, implying that the symmetry center is the optimal offsets where the LO leakage is maximally suppressed. We vary the LO-qubit detuning $\Delta f = f_{\rm LO} - f_{qubit}$ to show the robustness of the calibration scheme. It is found that as $|\Delta f|$ decreases, the regular egg-like pattern shrinks and butterfly-like pattern emerges from the boundary but the optimal offsets remain unchanged. This behavior aligns with our simulation results~\cite{sppm}. The excitation patterns exhibit rotational invariance due to their weak dependence on the angle $\theta=\arctan(Q_0/I_0)$, while the oscillating patterns depend on the amplitude of the drive field~\cite{sppm}. Similarly, the mirror sideband can be calibrated by tuning the qubit close to $f_{S}$. Scanning the correction parameter $c$ allows us to find the optimal value which minimizes qubit excitation, as shown in Fig.~\ref{fig:dline}(c). The envelope $A(t)$ used in experiments is a 10~$\mu$s square pulse whose amplitude is carefully minimized to a level just sufficient for driving the qubit, which also accounts for the fluctuations in the resulting patterns. The $|0\rangle$ population distribution exhibits centrosymmetric patterns at various sideband-qubit detunings $\Delta f = f_{S} - f_{qubit}$ because their interactions with the qubit are essentially equivalent, consistent with the simulation results~\cite{sppm}.

We employ a center-searching algorithm to identify the symmetry center corresponding to the optimal calibration parameters, ensuring robustness against a rugged parameter landscape. Our algorithm follows two steps. (i) Find the initial center of gravity (CoG), given by $\vec{G} = \sum \vec{g}_{xy} p_{xy}$, where $\vec{g}_{xy}=(x,y)$ represents the scanned parameter pair in vector form, $p_{xy}$ denotes the corresponding population, and the summation extends over all scanned parameters. (ii) Update the CoG iteratively by searching for a center with improved centrosymmetry until the result converges. The centrosymmetry of an arbitrary point $(x, y)$ is defined as
\begin{equation}
    \text{loss}(x, y) = \frac{1}{|R|} \sum_{(x^{\prime}, y^{\prime}) \in R} \left| p(x^{\prime}, y^{\prime}) - p(2x - x^{\prime}, 2y - y^{\prime}) \right|
    \label{eq:loss function}
\end{equation}
where \( R \) is the set of all scanned parameter pairs whose distance to $(x, y)$ is less than a given radius, which is empirically set to half of the scanned range. Note that if the population of the symmetric counterpart $p(2x - x^{\prime}, 2y - y^{\prime})$ is not measured in the experiments, the nearest available data is used as an estimation. In Fig.~\ref{fig:dline}(d), we use the data from the top right panel of Fig.~\ref{fig:dline}(b) as an example to visualize our algorithm. The initial CoG is marked as a red star in the left panel. The loss function naturally exhibits a global minimum whose coordinates correspond to the optimal parameters. For this instance, the algorithm converges within four iteration steps, which are visualized in the right panel. Our center-searching algorithm produces consistent results across all our experiments, demonstrating robust performance.

\begin{figure}
\includegraphics[width=0.5\textwidth]{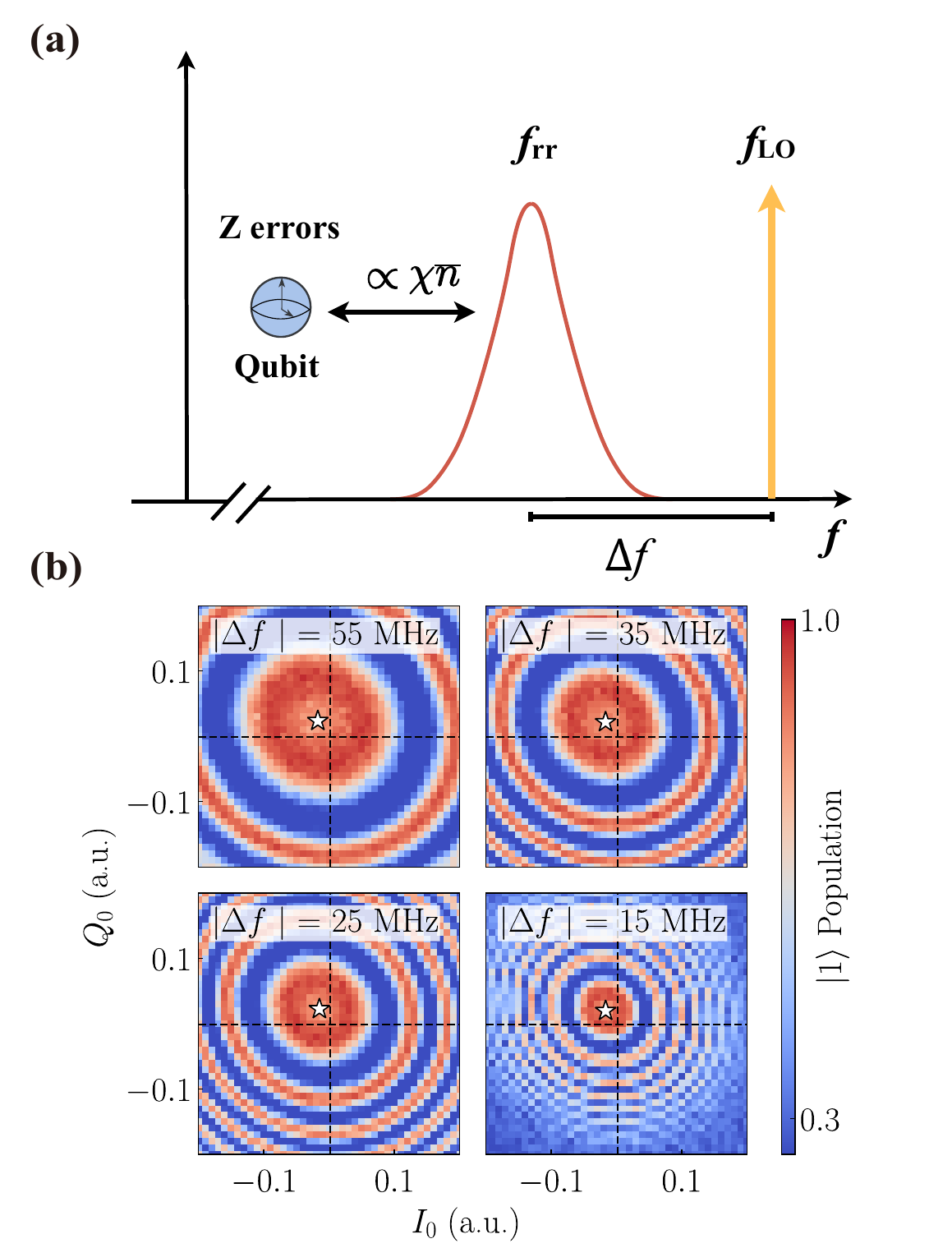}
\caption{\label{fig:rline} (a) LO leakage generates photons in qubit's readout resonator at the frequency $f_{\rm rr}$, which cause frequency shift of the qubit (Z errors), in spite of a finite LO-resonator detuning $\Delta f = f_{\rm LO} - f_{\rm rr}$. The frequency shift is proportional to the interaction strength between the qubit and the resonator $\chi \overline{n}$. We calibrate the measurement mixers by minimizing Z errors. (b) Oscillation patterns of $|1\rangle$ population as a function of mixer offsets at different detunings. The tendency of the change in the patterns is independent of the sign of $\Delta f$.}
\end{figure}

LO leakage of the mixers in the measurement line (measurement mixers) generates coherent photons in the readout resonators at a frequency $f_{\rm rr}$~\cite{PhysRevLett.120.260504}, even if the two signals are off-resonant. As illustrated in Fig.~\ref{fig:rline}(a), the leakage-induced photons cause measurable qubit frequency shift (Z errors) $\Delta _{qubit} \propto \chi \overline{n}$~\cite{sankthesis}, where $\overline{n}$ is the average photon number in the resonator, and $\chi$ is the dispersive shift of the resonator frequency. These leakage-induced Z errors provide a reliable metric for LO leakage calibration. With the qubit biased to its sweet spot (maximum frequency) to minimize flux noise, we employ Ramsey experiments as indicators of Z errors manifested as variations in $|1\rangle$ population. Centrosymmetry patterns are observed at various LO-resonator detuning, as shown in Fig.~\ref{fig:rline}(b), suggesting the feasibility of the center-searching algorithm introduced earlier. The leakage-induced Z errors are observed at a detuning as large as 55~MHz, indicating that our method is applicable considering limited tunability of readout resonator frequency. The observed patterns, resembling Newton's rings, qualitatively align with our theoretical model~\cite{sppm}, thereby validating our experimental approach. We focus on leakage calibration in this Letter as the image sideband signal of the measurement pulse exerts a much weaker effect due to its occurrence only during up-conversion. However, we propose that this method can be effectively extended to address sideband calibration, given the similar underlying principles.

\begin{figure}
\includegraphics[width=0.5\textwidth]{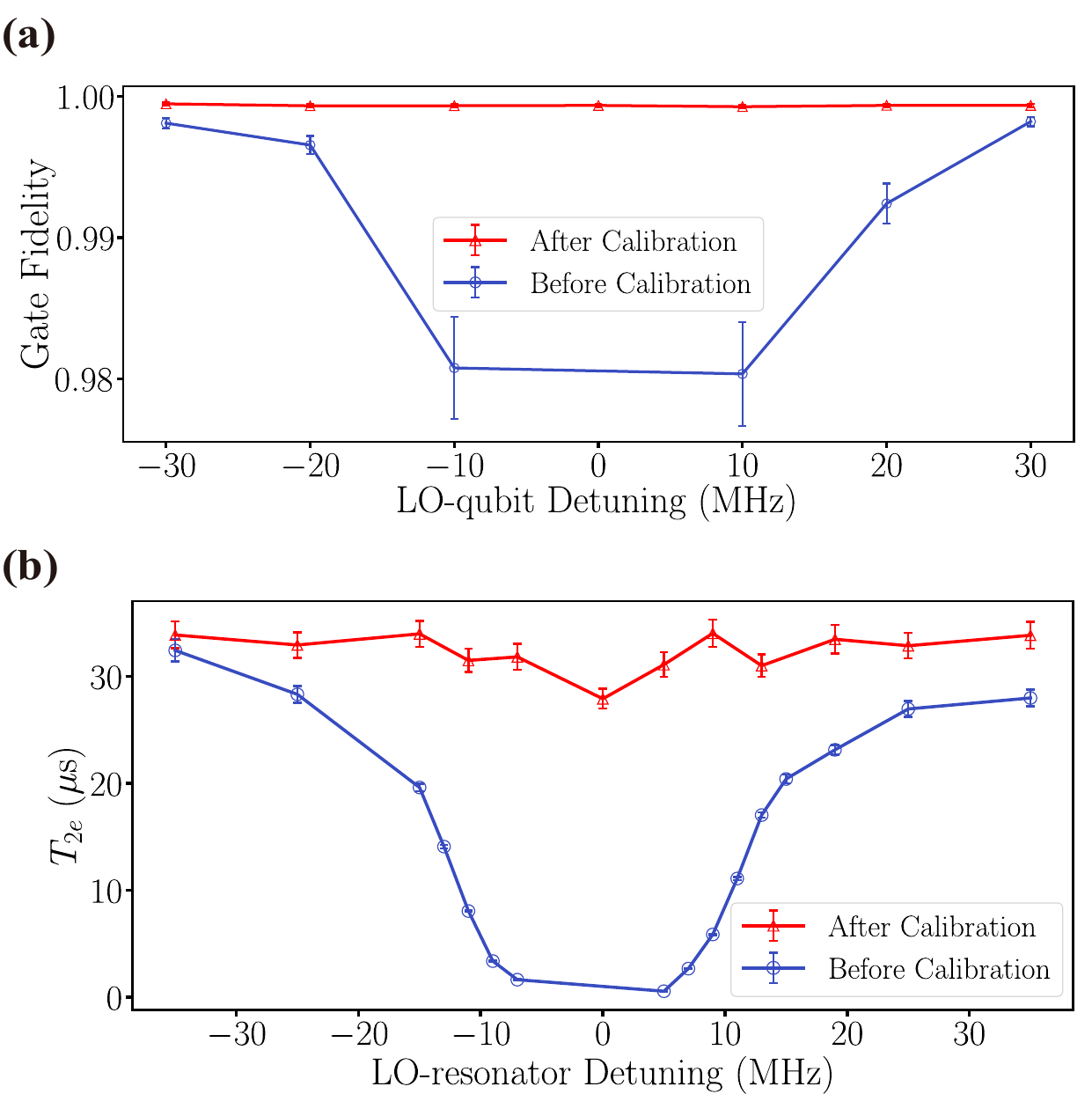}
\caption{\label{fig:RB} (a) Single-qubit gate fidelity before (blue) and after (red) drive mixer calibration shown across different LO-qubit detunings. The error bars for both curves are rescaled by a factor of 0.1 for better visualization. (b) Comparison of $T_{2e}$ before (blue) and after (red) calibrating the measurement mixer, plotted against LO-resonator detuning.}
\end{figure}

We evaluate our calibration method by benchmarking single-qubit gate and monitoring qubit coherence time. The single-qubit gate fidelity, assessed using randomized benchmarking~\cite{RB_PRA}, improves significantly after calibrating the drive mixer especially when the qubit frequency is close to $f_{\rm LO}$, as shown in Fig.~\ref{fig:RB}(a). We identified a critical `blind range' ($|\Delta f|<10$~MHz) where achieving high-fidelity control was unfeasible without proper calibration, highlighting the effectiveness of our method.

The calibration method for the measurement mixer is evaluated by monitoring qubit coherence time at various LO-resonator detuning. We measure qubit dephasing time $T_{2e}$ using spin-echo technique at the sweet point after calibrating the drive mixer, since this approach largely excludes flux noise and provides a more accurate evaluation of gate performance. The results, visualized in Fig.~\ref{fig:RB}(b), demonstrate that $T_{2e}$ declines to below 5~$\mu$s before calibration when the LO-resonator detuning $|\Delta f|<10$~MHz but exceeds 30~$\mu$s across arbitrary detunings after calibration. We also identify a `blind range' ($|\Delta f| < 5$~MHz) where values of $T_{2e}$ before calibration are absent due to the inability to perform dispersive readout caused by strong leakage.

In summary, we have demonstrated an \textit{in situ} method for calibrating IQ mixers using superconducting qubits. Our method does not require external instruments, thereby preserving the integrity of the closed control setup and reducing the burden of altering wiring configurations. The experimental sequences we employ are straightforward, and the resulting data can be efficiently processed using our center-searching algorithm, which optimizes calibration by locating the symmetry center of the qubit's response patterns. Our method significantly reduces control errors of single-qubit gates and minimizes qubit dephasing caused by mixer imperfections. Importantly, these improvements are effective across a wide range of detuning frequencies, ensuring robust performance even under suboptimal conditions, which is critical for scaled superconducting quantum computation.

\section*{Supplementary materials}
See supplementary materials for detailed results on theoretical model of qubit's response to imperfect mixer output.

\begin{acknowledgments}
This work was supported by the Science, Technology and Innovation Commission of Shenzhen Municipality (KQTD20210811090049034, RCBS20231211090824040, RCBS20231211090815032), the National Natural Science Foundation of China (12174178, 12204228, 12374474 and 123b2071), the Innovation Program for Quantum Science and Technology (2021ZD0301703), the Shenzhen-Hong Kong Cooperation Zone for Technology and Innovation (HZQB-KCZYB-2020050), and Guangdong Basic and Applied Basic Research Foundation (2024A1515011714, 2022A1515110615).
\end{acknowledgments}

\section*{Author Declarations}

\subsection*{Conflict of Interest}
The authors declare no competing interests.

\subsection*{Data Availability Statement}
The data that support the findings of this study are available from the corresponding author upon reasonable request.

\bibliography{main_ref}

\end{document}



\title{Supplementary material for \\ \textit{In situ} mixer calibration for superconducting quantum circuits}
\author{Nan Wu}
\affiliation{\SIQSE}\affiliation{\IQA}\affiliation{\GDKL}
\author{Jing Lin}
\affiliation{\SIQSE}\affiliation{\IQA}\affiliation{\GDKL}
\author{Changrong Xie}
\affiliation{\SIQSE}\affiliation{\IQA}\affiliation{\GDKL}
\author{Zechen Guo}
\affiliation{\SIQSE}\affiliation{\IQA}\affiliation{\GDKL}
\author{Wenhui Huang}
\affiliation{\SIQSE}\affiliation{\IQA}\affiliation{\GDKL}
\author{Libo Zhang}
\affiliation{\SIQSE}\affiliation{\IQA}\affiliation{\GDKL}
\author{Yuxuan Zhou}
\affiliation{\IQA}
\author{Xuandong Sun}
\affiliation{\SIQSE}\affiliation{\IQA}\affiliation{\GDKL}\affiliation{\DPHY}
\author{Jiawei Zhang}
\affiliation{\SIQSE}\affiliation{\IQA}\affiliation{\GDKL}
\author{Weijie Guo}
\affiliation{\IQA}
\author{Xiayu Linpeng}
\affiliation{\IQA}
\author{Song Liu}
\affiliation{\SIQSE}\affiliation{\IQA}\affiliation{\GDKL}\affiliation{\HFNL}
\author{Yang Liu}
\affiliation{\IQA}
\author{Wenhui Ren}
\affiliation{\IQA}
\author{Ziyu Tao}
\affiliation{\IQA}
\author{Ji Jiang}
\email{jiangj3@sustech.edu.cn}
\affiliation{\SIQSE}\affiliation{\IQA}\affiliation{\GDKL}
\author{Ji Chu}
\affiliation{\IQA}
\author{Jingjing Niu}
\affiliation{\IQA}\affiliation{\HFNL}
\author{Youpeng Zhong}
\affiliation{\SIQSE}\affiliation{\IQA}\affiliation{\GDKL}\affiliation{\HFNL}
\author{Dapeng Yu}
\affiliation{\SIQSE}\affiliation{\IQA}\affiliation{\GDKL}\affiliation{\DPHY}\affiliation{\HFNL}

\date{\today}

\maketitle
The supplementary material is organized into three sections: the theoretical model of the qubit's response to local oscillator (LO) leakage of drive mixer, the mirror sideband of drive mixer, and the LO leakage of measurement mixer. In the following analysis, we denote linear frequency as $f$ and angular frequency as $\omega$. For example, $f_{\rm LO}$ refers to frequency of LO as in the main text, while $\omega _{\rm LO}= 2 \pi f_{\rm LO}$ represents the corresponding angular frequency. The term `frequency' may indicate either angular frequency or linear frequency, depending on the context specified by the label, to avoid any ambiguity. Additionally, we set $\hbar = 1$ for simplicity.

\section{LO leakage of drive mixers}
\begin{figure*}
    \centering
    \includegraphics[width=1\textwidth]{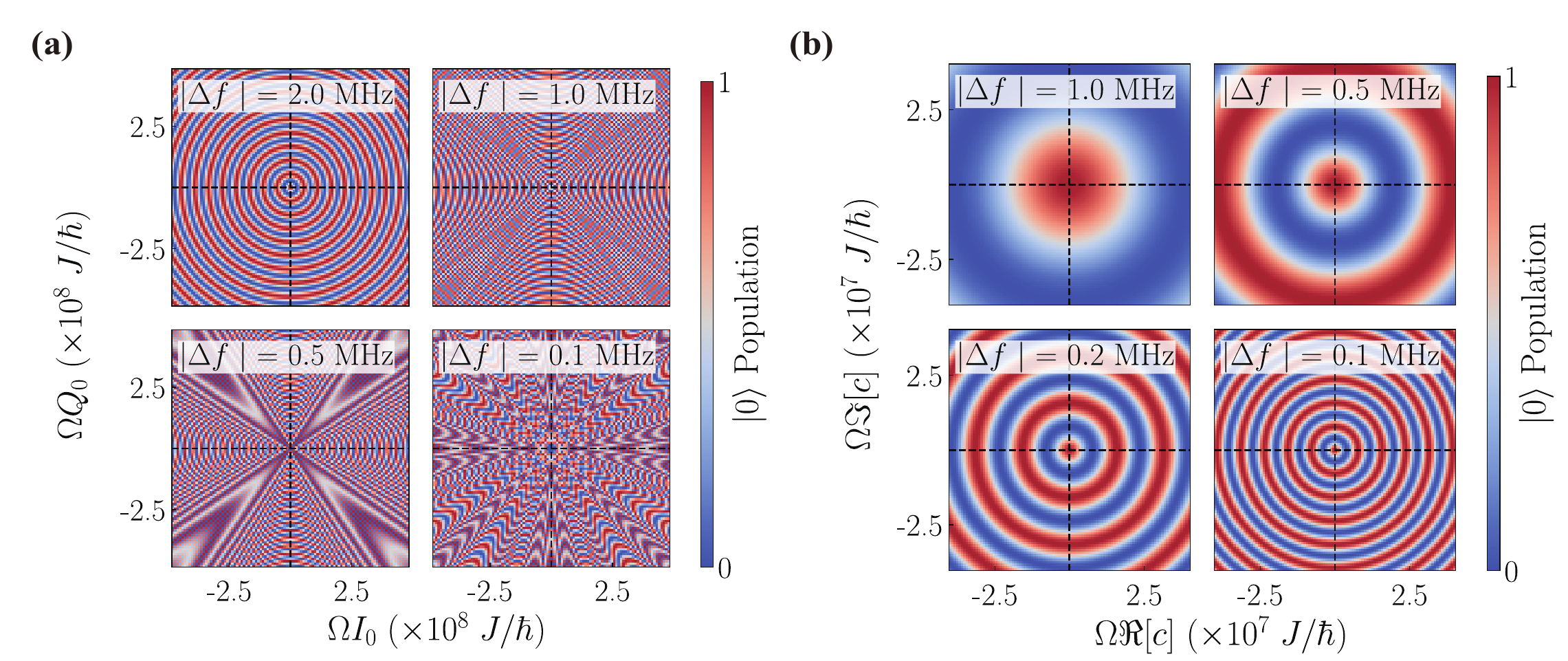}
    \caption{\label{fig:drive_simulations} (a) Ground state population of a qubit driven by the LO leakage microwave field as a function of the mixer offsets, which corresponds to the field amplitude. (b) Numerical results of excitation patterns driven by mirror sideband signals at different detunings.}
\end{figure*}

Given that the output signals generated by drive mixers directly affect qubits, we follow the approach discussed in circuit quantum electrodynamics~\cite{RevModPhys.93.025005} and analyze the perturbative component of the superconducting qubit's Hamiltonian in the Dirac picture. The interaction between a qubit and a drive field in rotating frame is
\begin{equation}
    \hat{H}_d = \Omega V_d (\cos{\omega_q t}\cdot\sigma_y - \sin{\omega_q t}\cdot\sigma_x),
    \label{qubit_drive_term}
\end{equation}
where $\sigma _x$ and $\sigma _y$ are the Pauli operator, $V_d=V_d(t)$ is the microwave drive field of LO leakage, and $\omega _q$ is the qubit frequency. $\Omega = (C_d/C_{\Sigma})Q_{zpf}$, where $C_d$ is the capacity of the drive line, $C_{\Sigma}$ is the total capacity of the drive line and the qubit, $Q_{zpf}=\sqrt{\hbar / 2Z}$ is the zero-point charge fluctuation and $Z$ is the impedance of the superconducting circuit to ground~\cite{engineerguide}. Following Eq.3-5 in the main text, in the absence of IF signals ($A(t)=0$), the RF output of drive mixer exhibits only the leakage signal
\begin{equation}
    \begin{aligned}
        V_d(t) &=  a I_0\cos \omega_{\rm LO} t - a Q_0\sin \omega_{\rm LO} t,
    \end{aligned}
    \label{eq:drive_leakage}
\end{equation}
where $a$ is the attenuation factor, and $I_0$ and $Q_0$ are the DC offsets of I and Q ports, respectively. For simplicity we assume $a=1$. Substitute Eq.~\ref{eq:drive_leakage} into Eq.\ref{qubit_drive_term} and integrate from $t=0$ to $t=t_0$, we have the propagator
\begin{equation}
    \begin{aligned}
        \hat{S}_d(t_0, 0) &= \exp (\frac{1}{i}\int_0^{t_0} \mathrm{d}t \hat{H}_d) \\
        &= \exp (\frac{\Omega V_0}{2i}
        \{[\frac{\sin (\delta\omega t+\theta)}{\delta\omega} + \frac{\sin (\omega_\Sigma t+\theta)}{\omega_\Sigma}]\sigma_y
        - [\frac{\cos (\delta\omega t+\theta)}{\delta\omega} - \frac{\cos (\omega_\Sigma t+\theta)}{\omega_\Sigma}]\sigma_x
        \}|_0^{t_0}),
    \end{aligned}
    \label{eq:propagator}
\end{equation}
where $\theta = \arctan (Q_0/I_0)$, $V_0 = \sqrt{I_0^2 + Q_0^2}$, $\omega_\Sigma = \omega_{\rm LO}+\omega_q$ and $\delta\omega = \omega_{\rm LO}-\omega_q \neq 0$. Note that the time-oscillating terms vanish by integrating, thus do not affect the symmetry pattern we are interested in. By introducing $\hat{S}_d(t_0, 0) = \hat{U}(t_0)\hat{U}^\dagger(0)$ and $z=z_0 e^{i \eta}$, we may write the operator $U$ as
\begin{equation}
    \hat{U} = \exp (\frac{1}{i} \begin{bmatrix}0 & z^\dagger \\ z & 0 \end{bmatrix})
    = \begin{bmatrix}
        \cos z_0 & -ie^{-i\eta} \sin z_0 \\
        -ie^{i\eta} \sin z_0 & cos z_0
    \end{bmatrix}.
    \label{eq:matrix_exp}
\end{equation}
Therefore, the ground state population of a leakage-driven qubit at $t=t_0$ is $P(|0\rangle) = \cos ^2 z_0$, where
\begin{equation}
    \begin{aligned}
        \frac{2z_0}{\Omega V_0} = \sqrt{2\cdot(\frac{1}{\delta\omega^2}+\frac{1}{\omega_\Sigma^2}-\frac{\cos (2\omega_{\rm LO} t_0 + 2\theta)}{\delta \omega \cdot \omega_\Sigma})},
        \label{eq:qb_z0}
    \end{aligned}
\end{equation}
and
\begin{equation}
    \frac{\partial}{\partial \theta} (\frac{2z_0}{\Omega V_0}) = \frac{\sqrt{2}\sin (2\omega_{\rm LO}t+2\theta)}{\omega_\Sigma \sqrt{1+(\frac{\delta\omega}{\omega_\Sigma})^2-\frac{\delta\omega}{\omega_\Sigma}\cos (2\omega_{\rm LO}t+2\theta)}}.
\end{equation}
In experiments, the order of magnitude of $\omega _{\Sigma}$ is 10~GHz, $\delta \omega \sim 1$~MHz and $\Omega V_0$ is 10~MHz, indicating that $\partial _{\theta} P(|0\rangle) \sim 0$, which accounts for the rotational symmetry observed in experiments. On the other hand, since the frequency of the population oscillation is proportional to the amplitude of the drive field $V_0$, we observe complex patterns at small detunings. Numerical results performed with QuTiP~\cite{johansson2012qutip,JOHANSSON20131234} show centrosymmetric patterns at various LO-qubit detunings, as shown in Fig.~\ref{fig:drive_simulations}(a). The egg-like patterns observed in experiments, i.e., the upper-left panel of Fig.2(a) in the main text, corresponds to the central part of the patterns assembling Newton's ring shown in the upper-left panel of Fig.~\ref{fig:drive_simulations}(a). Our simulation results also reproduce the complex butterfly-like patterns at small detunings, validating our experiment approach.

\section{Mirror sideband of drive mixers}
Assume LO leakage has been compensated before calibrating mirror sideband. Following Eq.1 and Eq.2 in the main text, with the mirror sideband correction parameter being a complex parameter $c$, drive field of the corrected mirror sideband is
\begin{equation}
    V_d(t) = A(t)\{\Re[c] \cos[( \omega_{\rm LO} - \omega_{\rm IF})t] - \Im[c] \sin[(\omega_{\rm LO} - \omega_{\rm IF})t]\} \label{eq:drive_sideband}.
\end{equation}
In experiments, frequency of the RF signals generated by the mixer is $\omega_{\rm LO} + \omega_{\rm IF}$, meanwhile the qubit frequency is biased to the frequency of the mirror sideband $\omega_{\rm LO} - \omega_{\rm IF}$. The sideband-qubit detuning is $\delta \omega = \omega_{\rm LO} - \omega_{\rm IF} - \omega _q$. $A(t)=A$ is a constant since the envelope is a square pulse, according to the description in the main text. In simulation, we assume the amplitude of the target frequency is small, therefore the effect of the off-resonant frequency $\omega _{\rm LO} + \omega _{\rm IF}$ is negligible. The drive field is in principle identical to the leakage model Eq.~\ref{eq:drive_leakage}. Following similar steps, the ground state population of the qubit can be written as $P(|0\rangle) = \cos ^2 z_0$, while $z_0$ has a similar form to Eq.~\ref{eq:qb_z0}:
\begin{equation}
\begin{aligned}
    \frac{4z_0}{\Omega}
    = A |c| \cdot \sqrt{ \frac{1}{(2\omega_q+\delta\omega)^2} + \frac{1}{\delta\omega^2} + \frac{2\cos (2\omega_qt+2\delta\omega t+2\phi)}{\delta\omega(2\omega_q+\delta\omega)} }.
\end{aligned}
\end{equation}
Due to this similarity, patterns of the ground state population driven by the mirror sideband signal, in principle, exhibit the same behavior as those driven by LO leakage, as supported by our simulation results shown in Fig.~\ref{fig:drive_simulations}. However, complex patterns are not observed since the drive field is intentionally weakened, as mentioned in the main text.

\begin{figure*}
    \centering
    \includegraphics[width=1\textwidth]{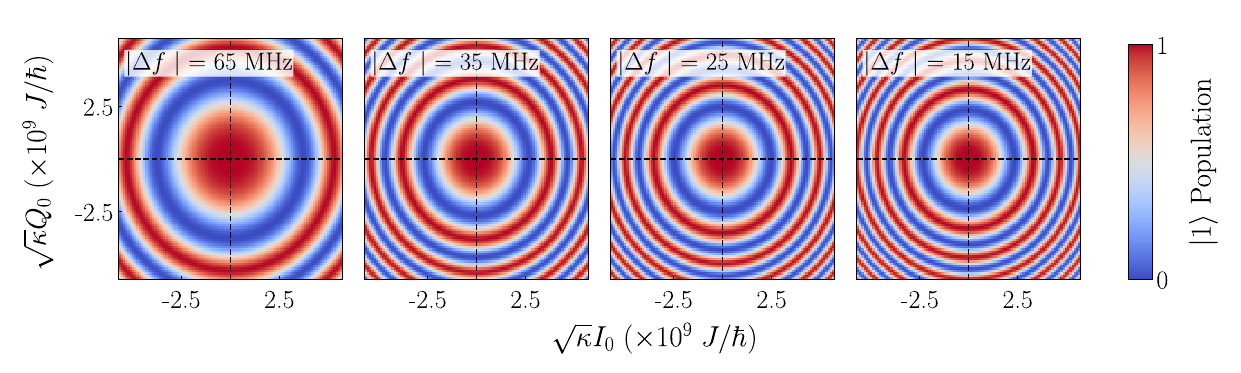}
    \caption{\label{fig:read_simulations} Excitation patterns due to LO leakage of measurement mixer given by simulations at different LO-resonator detunings.}
\end{figure*}

\section{Read-in Mixer leakage induced qubit detuning}
Here, we delve into how off-resonant resonator driving leads to qubit detuning, while neglecting the photon-induced qubit dephasing. 
We consider a resonator with frequency $\omega_r$ coupled with a qubit with frequency $\omega _q$ is driven by a microwave field $A(t)e^{i\omega_{\rm LO} t}$ due to LO leakage of measurement mixer. Following the Jaynes-Cumming model~\cite{engineerguide}, the entire system is divided into four parts: the qubit, the resonator, interaction between the qubit and the resonator, and the microwave drive field:
\begin{equation}
\begin{aligned}
    H_{q} &= \frac{\omega_q}{2} \sigma_z \\
    H_{r} &= \omega_r (a^{\dagger} a + \frac{1}{2}) \\
    H_{int} &= g(\sigma_{+} a + \sigma_{-} a^{\dagger}) \\
    H_{d} &= \mu[i\sqrt{\kappa}A(t)a^\dagger \exp (i\omega_{\rm LO} t)+h.c.],
\end{aligned}
\end{equation}
where $\Gamma$ is the linewidth of the resonator and $\mu = 1/[1+(\frac{\omega_{\rm LO}-\omega_{r}}{\Gamma})^2]$ is the coupling strength between the cavity and the transmission line. Let $V_d (t) = \Re [A(t)e^{i\omega_{\rm LO} t}]$ and define
\begin{equation}
\begin{aligned}
    U_q &= e^{-i(\omega_{\rm LO} t) \sigma_+ \sigma_-} \\
    U_r &= e^{-i(\omega_{\rm LO} t) a^\dagger a} \\
    U &= U_q U_r.
\end{aligned}
\end{equation}
Applying the similarity transformation to the system Hamiltonian leads to a time-independent form~\cite{RevModPhys.93.025005}. We numerically solve the state evolution without considering the resonator dampling:
\begin{equation}
    \ket{\psi (t)} = U^\dagger (t_2) e^{-iH'(t_2-t_1)} U (t_1) \ket{\psi}_{t=0}.
\end{equation}
The simulation results without photon-induced dephasing is given in Fig.~\ref{fig:read_simulations} exhibit behavior identical to that observed in the experiments discussed in the main text, thereby validating our experimental approach.

\bibliography{supple_ref}